\documentclass[twocolumn, superscriptaddress]{revtex4-2}
\usepackage{graphicx}
\usepackage{physics}
\usepackage{amssymb}
\usepackage{amsmath}
\usepackage{color}
\usepackage{graphicx}
\usepackage{dcolumn}
\usepackage{amsfonts}
\usepackage{bm}
\usepackage{natbib}
\usepackage{tikz}
\usepackage[colorlinks=true,citecolor=blue]{hyperref}

\begin{document}
\title{High-efficiency microwave photodetection by cavity coupled double dots with single cavity-photon sensitivity}
\author{Subhomoy Haldar}
\email{subhomoy.haldar@ftf.lth.se}
 \address{NanoLund and Solid State Physics, Lund University, Box 118, 22100 Lund, Sweden}
 \address{Department of Physics, Indian Institute of Technology Kanpur, Kanpur, Uttar Pradesh 208016, India}
 \author{Harald Havir}
 \address{NanoLund and Solid State Physics, Lund University, Box 118, 22100 Lund, Sweden}
 \author{Waqar Khan}
 \address{NanoLund and Solid State Physics, Lund University, Box 118, 22100 Lund, Sweden}
\address{Presently at Low Noise Factory AB, 41263 Göteborg, Sweden.}
 \author{Drilon Zenelaj }
 \address{NanoLund and Mathematical Physics, Lund University, Box 118, 22100 Lund, Sweden}
 \author{Patrick P. Potts}
 \address{Department of Physics and Swiss Nanoscience Institute, University of Basel, Klingelbergstrasse 82, 4056 Basel, Switzerland}
 \author{Sebastian Lehmann}
 \address{NanoLund and Solid State Physics, Lund University, Box 118, 22100 Lund, Sweden}
\author{Kimberly A. Dick}
 \address{NanoLund and Solid State Physics, Lund University, Box 118, 22100 Lund, Sweden}
 \address{Center for Analysis and Synthesis, Lund University, Box 124, 22100 Lund, Sweden}
  \author{Peter Samuelsson}
 \address{NanoLund and Mathematical Physics, Lund University, Box 118, 22100 Lund, Sweden}
 \author{Ville F. Maisi}
 \email{ville.maisi@ftf.lth.se}
 \address{NanoLund and Solid State Physics, Lund University, Box 118, 22100 Lund, Sweden}
\date{\today}
\begin{abstract}

We present a superconducting cavity-coupled double quantum dot (DQD) photodiode that achieves a maximum photon-to-electron conversion efficiency of 25\% in the microwave domain. With a higher-quality-factor cavity and improved device design to prevent photon leakages through unwanted pathways, our device measures microwave signals down to 100 aW power level and achieves sensitivity to probe microwave signals with one photon at a time in the cavity. We analyze the photodiode operation using Jaynes-Cummings input-output theory, identifying the key improvements of stronger cavity-DQD coupling needed to achieve near-unity photodetection efficiency. The results presented in this work represent a crucial advancement toward near-unity microwave photodetection efficiency with single cavity-photon sensitivity. 

\end{abstract}
\maketitle

\section{Introduction}

The investigation of microwave photons interacting with solid-state quantum systems plays a central role in advancing quantum information processing, quantum communication, and precision measurements for quantum computing~\cite{gu2017, blais2020, brecht2016, heinsoo2018}. In this domain, an intriguing approach involves integrating a superconducting cavity with semiconductor quantum dots (QDs)~\cite{chatterjee2021, burkard2020, ares2013}. The QDs, characterized by their discrete energy levels, offer a unique platform for exploring fundamental quantum optics effects while simultaneously allowing electron transport and realizing key elements for quantum technology~\cite{garcia2021, uppu2021, havir2023}. Here, a double quantum dot (DQD) system is particularly important because of its gate voltage-controlled tunnel-coupled energy levels~\cite{burkard2020, chatterjee2021}. Extensive research has been performed to investigate the interaction of microwave photons with DQDs~\cite{kouwenhoven1994, vanderWiel2002, liu2015, Gullans2015, stockklauser2015, ghirri2020}. This includes studies of photon-assisted electron tunneling for applications related to microwave photodetection~\cite{khan2021, ghirri2020, wong2017} and photon emission during electron tunneling demonstrating on-chip maser operation~\cite{liu2015, liu2017}. While superconducting circuits have been used to demonstrate near-unity photodetection efficiency in the microwave domain~\cite{opremcak2021, stanisavljevic2023, wang2023single}, such advancements with semiconductor QDs are encouraging, especially due to their gate voltage-controlled discrete energy levels that offer energy selectivity for the incident photons and integration into existing semiconductor technology~\cite{petersson2010, cao2016}. 
Another promising approach for photodetection involves superconducting hot-electron nano-bolometers, which provide higher bandwidth and do not require spectral tuning, unlike semiconductor QD-based systems~\cite{karasik2011}. However, these bolometer detectors typically operate in the infrared and THz regimes, where photon energies are more than an order of magnitude higher than the 4-8 GHz frequency microwave photons. Therefore, the remarkable advancements made in these fields are not yet directly relevant to the present study focusing on low-frequency GHz photons.

\begin{figure*}
\includegraphics[width=6.2in]{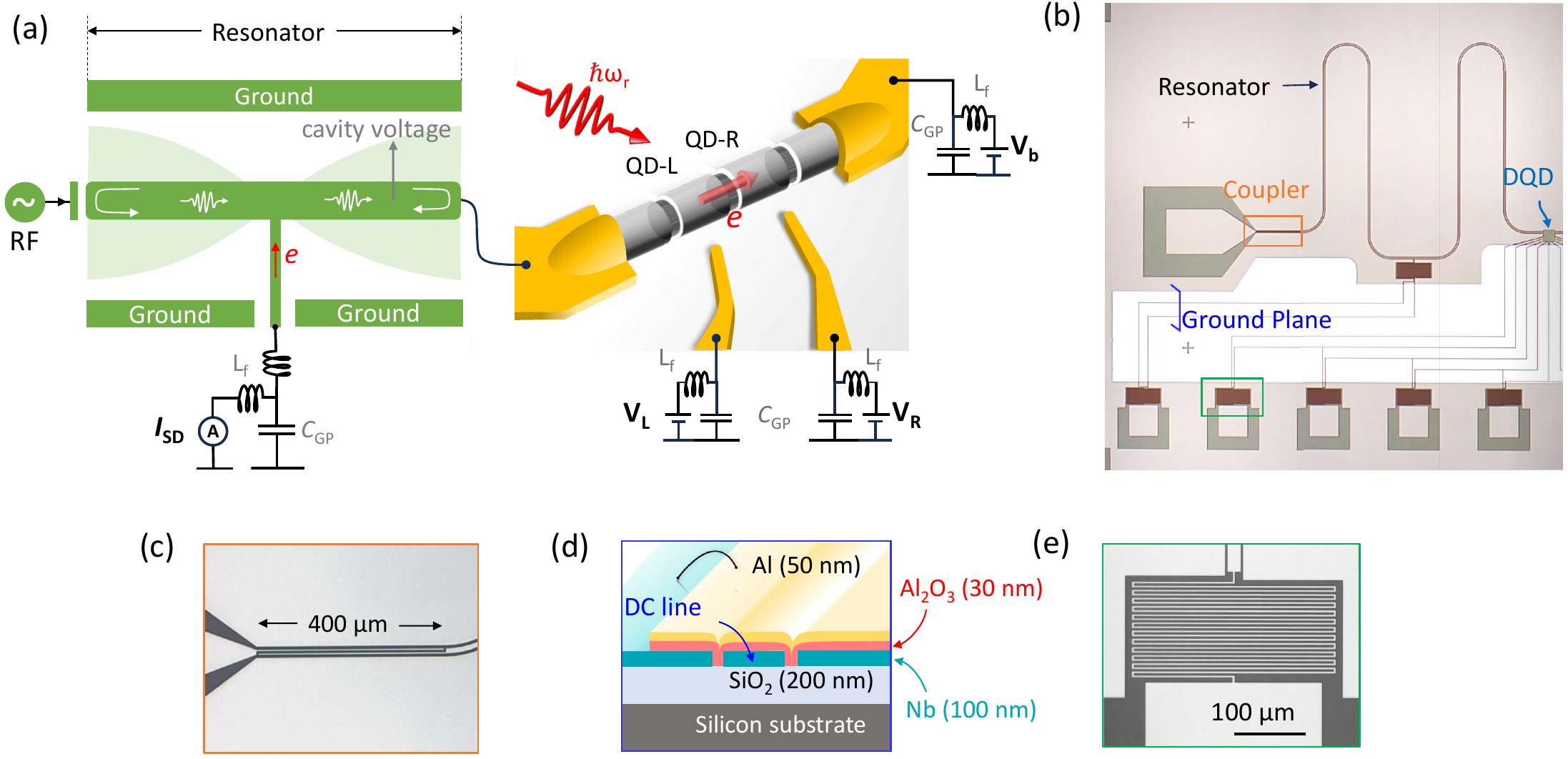}
\caption{\label{fig1}(a) Schematic illustration of the hybrid device, comprising a semiconductor nanowire double quantum dot coupled to a coplanar waveguide cavity. The photocurrent of the device is measured by utilizing the cavity voltage node at the mid-point of the cavity. (b) Optical microscope image of the chip showing the one-port cavity and indicating the input port, ground plane on the DC lines, and location of the DQD. (c) Input coupler of the cavity. (d) Schematic cross-sectional view showing the layered structure of the ground plane capacitor on the DC lines and (e) an inductive filter integrated into a DC line.}
\end{figure*}

Recent experiments by Khan \textit{et al.}~\cite{khan2021} demonstrated a cavity-coupled semiconductor DQD as a microwave photodetector with 6\% photo-conversion efficiency. As compared to the results presented in Ref.~\citealp{khan2021}, in this work and in Refs.~\citealp{haldar2022} and \citealp{haldar2023}, we improved (i) the internal losses of the resonator by replacing the resonator material from Al to Nb, (ii) minimized photon leakages by adding large capacitive filters on the dc lines and (iii) lowered the coupling losses using one-port resonator geometry instead of the two-port. While Ref.~\citealp{haldar2022} focused on studying the energetics of microwave absorption during photon-assisted electron tunneling and Ref.~\citealp{haldar2023} the energy conversion efficiency, here we investigate the improvements to the photodetection efficiency-- thanks to these aspects.

We demonstrate a superconductor-semiconductor hybrid device achieving a maximum photodetection efficiency of 25\% operating down to 100~aW input power level. We note that for studies of photon statistics and quantum sensing, it is crucial that (i) the detector achieves near-unity efficiency and (ii) can be operated with one (or fewer) photons in the cavity~\cite{fox2006quantum, hadfield2009}. Our work advances these goals by showing photodiode operation with a single cavity photon. We analyze the device operation using Jaynes-Cummings input-output theory and suggest future improvements to achieve near-unity photodetection efficiency by increasing the cavity-DQD coupling strength, potentially by introducing a high-impedance resonator~\cite{ranni2023, stockklauser2017}. We also investigate the dynamics of microwave-photon-assisted electronic transport in an asymmetrically tunnel-coupled DQD photodiode and characterize the performance parameters of our device.

\section{Device and Experimental Details}\label{sec2}

Our hybrid device consists of a semiconductor nanowire DQD, dipole-coupled to a superconducting coplanar waveguide (CPW) cavity, Fig.~\ref{fig1}(a)~\cite{khan2021, haldar2022, haldar2023}.  We use a polytype InAs nanowire with zinc blende and wurtzite crystal structures, grown using metal-organic vapor phase epitaxy~\cite{lehmann2013}. The electrons are confined within zinc blende islands due to a conduction band offset of approximately 120~meV between the two crystal phases~\cite{thelander2011, barker2019}. The nanowire, typically 100~nm in diameter, consists of 100~nm long zinc blende sections forming the dots and 20~nm long wurtzite barriers. The plunger gate voltages $V_\text{L}$ and $V_\text{R}$ tune the energy levels $\varepsilon_\text{L}$ and $\varepsilon_\text{R}$, and change the carrier occupancies \textit{N} and \textit{M} of the left and right dots, respectively. The one-port half-wavelength ($\lambda/2$) CPW cavity, made of a 100~nm thick sputtered Nb film, Fig.~\ref{fig1}(b), directly connects to the drain (D) lead of the DQD. The DQD is located at the voltage anti-node of the cavity mode, as depicted in Fig.~\ref{fig1}(a). The other end of the cavity connects to a microwave input port via a 400~$\mu$m long finger capacitor, Fig.~\ref{fig1}(c). The cavity has a characteristic resonance at $\omega_{r}/2\pi$ = 6.716~GHz. The measurements are performed in a dilution refrigerator at an electronic temperature of $T_e\approx 60$~mK. The same device has been used previously in Refs.~\cite{haldar2023, haldar2022}. 

In cavity-coupled QD devices, undesired photon leakages through the DC gate lines often lead to a low-quality factor of the cavity~\cite{khan2021, frey2012, petersson2012}. A plausible approach to mitigate such effects is to introduce on-chip low-pass capacitive and inductive filtering~\cite{mi2017, collard2020, zhang2024}. In our device, we address this by depositing a 30~nm thick Al$_2$O$_3$ layer followed by a 50~nm thick Al layer on the DC lines. See the bright-shaded area that appears in Fig.~\ref{fig1}(b) and the schematic cross-sectional view in Fig.~\ref{fig1}(d). The Al layer is then wire-bonded to the ground plane of the device. With this design, the capacitance of a DC line to the ground increases to $C_\mathrm{GP}$ $\approx$ 600~pF. Thus, the impedance of the DC line, for a high-frequency signal, reduces to $Z_\mathrm{line} \approx$ 40~m$\Omega$, which is more than three orders of magnitude smaller than the cavity impedance $Z_\mathrm{r}\approx$ 58~$\Omega$. Under this condition, only $2Z_\mathrm{line}/(Z_\mathrm{r}+Z_\mathrm{line}) \lesssim$ 0.1~\% of the microwave power may transmit into a single DC line. Additionally, we have integrated on-chip inductive ($L_f$) filters, Fig.~\ref{fig1}(e). These modifications allow us to obtain a maximum internal quality factor of $Q_\text{int} = \omega_r/\kappa_\text{int}=$ 9200 for a similar device with the DQD tuned to a very-small tunnel coupling regime. Here, $\omega_r$ is the cavity resonance frequency, and $\kappa_\text{int}$ is the internal losses of the cavity. However, for our current device, we have $Q_\text{int}$ = 5150 with the DQD tuned to a Coulomb-blockade regime.

\begin{figure}
\includegraphics[width=3.45in]{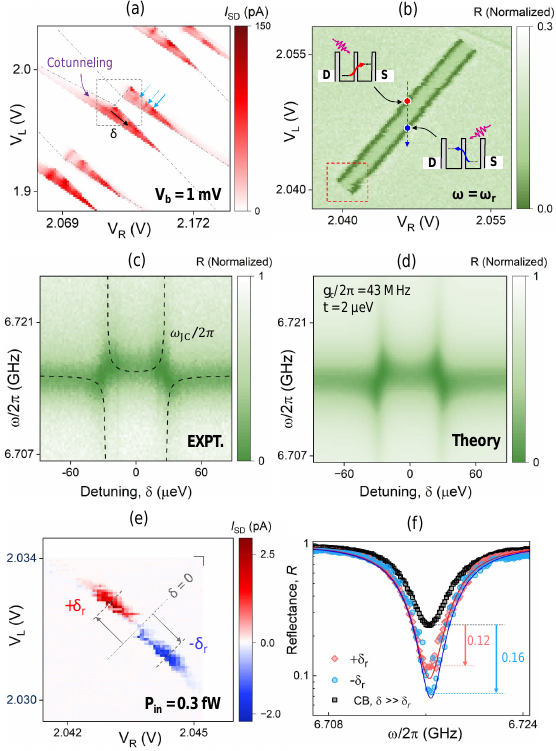}
\caption{\label{fig2}(a) Charge stability diagram of the DQD showing finite bias triangles (FBTs) at charge triple points with $V_b$ = 1~mV and without an RF drive. (b) Measured RF reflectance $R$ with $\omega=\omega_r$ as a function of plunger gate voltages $V_\text{L}$ and $V_\text{R}$ with $V_b$ = 0 and $P_\text{in}$ = 1~fW. Two distinct lines appear due to the DQD absorption at $\pm\delta_r$. (c) The measured RF spectra across the interdot charge transfer line and (d) theoretically fitted RF spectra using the Jaynes-Cummings model. The dashed lines in panel (c) show the lowest two transition energies of the Jaynes-Cummings Hamiltonian, Eq.~(\ref{Eq:dispersive}). (e) Photocurrent measurement across the plunger gate space with $P_\text{in} = 0.3$~fW and drive frequency 6.716~GHz. (f) Measured reflectance as a function of drive frequency with the DQD tuned to photodetection points at $\pm\delta_r$ at the charge triple points with finite photocurrent and the Coulomb blockade regime $\delta\gg\delta_r$ when $I_\text{SD}$ = 0, showing insights into cavity photon dissipation due to DQD absorption.}% 
\end{figure}

\section{Results}

To characterize the transport properties of the DQD, we begin by measuring the charge stability diagram around (\textit{N}, \textit{M}) $\leftrightarrow$ (\textit{N}+1, \textit{M}+1) charge occupations with a bias voltage of $V_b$ = 1~mV, Fig.~\ref{fig2}(a). The applied bias enables electron tunneling across the DQD, forming finite bias triangles (FBTs) at the triple points~\cite{liu2014, hartke2018, barker2019}. The black arrow in Fig.~\ref{fig2}(a) indicates the detuning ($\delta$) axis. The level detuning, $\delta=\varepsilon_\text{R}-\varepsilon_\text{L}$, tunable by the gate voltages, determines the energy gap $E_q=(\delta_r^2+4t^2)^{1/2}$ of the hybridized DQD levels $\ket{g}$ and $\ket{e}$~\cite{frey2012}, where $t$ denotes the tunnel coupling between the dot levels. Here, $\ket{g}$ and $\ket{e}$ denote the hybridized ground and excited states of the DQD, respectively. The FBTs exhibit distinct features of charge tunneling via the excited states of the QDs, as indicated by blue arrows in Fig.~\ref{fig2}(a). We also observe co-tunneling lines appearing along one edge of the hexagonal stability pattern~\cite{vanderWiel2002, barker2019}.  

Next, we measure the RF reflectance ($R$) with $\omega=\omega_r$ across the plunger gate space within the dashed square outlined in Fig.~\ref{fig2}(a). The measured result with input power $P_\text{in}$ = 1~fW is shown in Fig.~\ref{fig2}(b). We observe two distinct absorption lines at the inter-dot charge transfer region with $R\approx 0$. As shown in previous studies~\cite{frey2012, khan2021, wong2017}, photon-assisted tunneling of an electron is energetically permissible when the DQD energy gap $E_q$ matches the photon energy $\hbar\omega_r$. Therefore, the DQD at $\delta=\pm\delta_r$ introduces an additional loss channel to the cavity photon mode, resulting in the observation of two absorption lines in Fig.~\ref{fig2}(b). Further, to realize the cavity-DQD interactions, in Fig.~\ref{fig2}(c), we present the measured reflectance $R$ as a function of drive frequency $\omega/2\pi$ across the charge transfer line along the blue-dashed arrow in Fig.~\ref{fig2}(b). The corresponding theory result is shown in Fig.~\ref{fig2}(d), which we discuss in Section~\ref{deviceparams}.

\begin{figure}[t]
\includegraphics[width=3.1in]{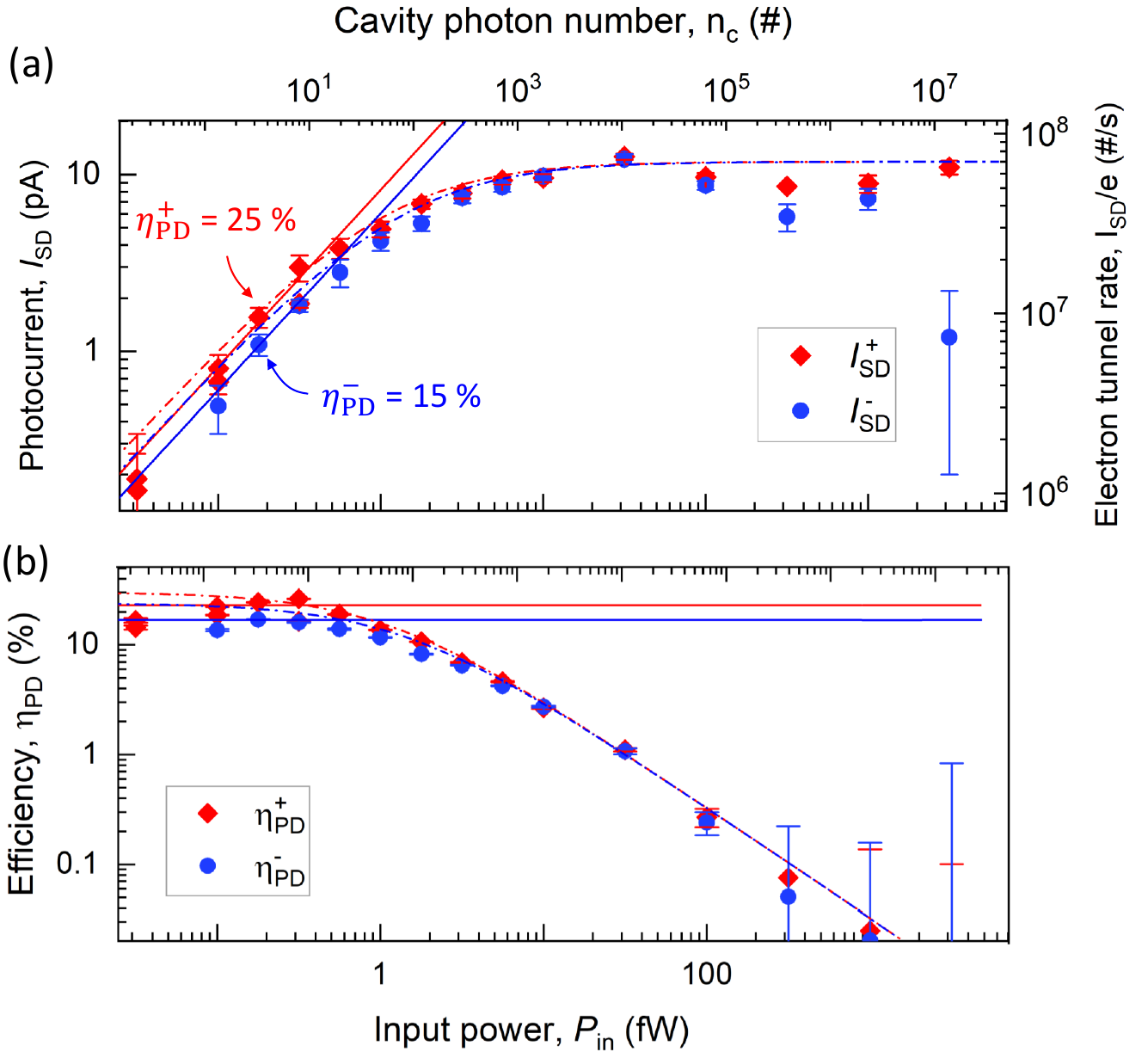}
\caption{\label{fig3}(a) Photocurrent recorded at $\pm\delta_r$ and (b) corresponding photodetection efficiency as a function of input microwave power, $P_\text{in}$. The electron tunneling rate $I_\text{SD}/e$ contributing to the measured signal is also shown on the right-hand side of panel a. The device achieves a maximum photoconversion efficiency of 25\%. The solid (dashed-dotted) lines show the theoretically fitted results in the low (high) power regime. }
\end{figure}

We now turn to measure the photocurrent $I_\text{SD}$ as a function of $V_\text{L}$ and $V_\text{R}$ with input power $P_\text{in}$ = 0.3~fW and bias $V_b$ = 0. Figure~\ref{fig2}(e) shows the measured result. We observe two distinct resonant features of positive ($I_\text{SD}^+$) and negative ($I_\text{SD}^-$) photocurrent at $+\delta_{r}$ and $-\delta_{r}$, respectively. We also observe a factor of 1.5 higher photocurrent at the positive detuning than at the other operation point. Such effects can be attributed to the asymmetric tunnel rates at the left ($\Gamma_{\text{L}}$) and right ($\Gamma_{\text{R}}$) QD-lead barriers of the DQD~\cite{haldar2023}. To probe the corresponding cavity photon dissipation due to the DQD absorption ($\kappa_\text{DQD}$), we record the reflectance spectra when operating the device at $\pm\delta_r$ with photocurrent $I_\text{SD}\neq 0$ and in the Coulomb blockade (CB) regime with $\delta\gg \delta_r$ and $I_\text{SD} =0$. Figure~\ref{fig2}(f) shows the measured results. The measured $R$$(\omega=\omega_r)$ changes by $\Delta R^+$ = -0.12 ($\Delta R^-$ = -0.16) when operating the device at $+\delta_r$ (-$\delta_r$). Note that the observed $I_\text{SD}^+>I_\text{SD}^-$ in Fig.~\ref{fig2}(e) would suggest a larger absorption at $+\delta_r$. However, contrary to this expectation, Fig.~\ref{fig2}(f) shows an opposite behavior: a lower photon dissipation at $+\delta_r$ than at $-\delta_r$. This indicates the role of competing processes contributing to the photodiode operation, which we discuss in detail below.

Next, we tune the DQD level detuning to $\pm\delta_r$ at the charge triple point and measure $I_\text{SD}$ as a function of $P_\text{in}$, Fig.~\ref{fig3}(a). The corresponding photodetection efficiency, $\eta_\text{PD} = I_\text{SD}\hbar\omega_r/eP_\text{in}$, at the two operation points is shown in Fig.~\ref{fig3}(b). Note that the electron tunnel rate shown on the right-hand side in Fig.~\ref{fig3}(a) demonstrates that the photon-to-electron conversion takes place at the MHz-level rates. The $I_\text{SD}$ in the low-drive limit, $P_\text{in}<$ 1~fW, increases linearly with $P_\mathrm{in}$. In this linear response regime, we obtain the photon-to-electron conversion efficiency, $\eta_\mathrm{PD}^-$ = 15\:$\pm$\:2 \% at $-\delta_r$. At $+\delta_r$, the photoconversion efficiency becomes $\eta_\text{PD}^+$ = 25\:$\pm$\:2 \%, which surpasses recent results for a similar cavity-coupled DQD device of Ref.~\citealp{khan2021} by fourfold and outperforms the results of Ref.~\citealp{haldar2023} by a factor of two. Another crucial advancement here is the ability to operate the device with an input power of $100$~aW, which is at least an order of magnitude lower than the power levels at which previous devices were operated~\cite{khan2021, haldar2023, cornia2023}. We note that at $P_\text{in}$ = 100~aW, the number of photons stored in the cavity reaches $n_c =2C_rQ_\text{L}^2Z_rP_\text{in}/Q_\text{ex}\hbar\omega_r$ = 1.35~\cite{haldar2022}, where $Q_\text{L} = 1300 $ ($Q_\text{ex} = 1720$) is the loaded (external) quality factor and $C_r=540$~fF is the total capacitance of the cavity. The cavity photon numbers $n_c$ for different input powers are also shown in Fig.~\ref{fig3}. Increasing the input power $P_\text{in}>$ 30~fW decreases the photo-response, Figs.~\ref{fig3}(a-b). In this regime, multiple-photon absorption processes typically contribute to photon-assisted tunneling, where the cavity-photon field modulates the inter-dot tunneling rate~\cite{tien1963, oosterkamp1998}. Investigation of such multi-photon processes is beyond the scope of our present discussion. 

\section{Device Parameters}\label{deviceparams}

We now turn to fit the measured results of Fig.~\ref{fig2} using the Jaynes-Cummings input-output theory model~\cite{frey2012, zenelaj2022, haldar2023, khan2021, wong2017} to obtain the device parameters. We begin by fitting the interaction of the cavity photons and the DQD of Fig.~\ref{fig2}(c), which can be described by~\cite{ranni2023},

\begin{equation}\label{eqS1}
    |R(\omega)|=|\kappa_c A(\omega)-1|^2,
\end{equation}
with
\begin{equation}\label{eqS2}
        A(\omega)=\frac{\Tilde{\Gamma}_0/2-i(\omega-\omega_q)}{[\kappa/2-i(\omega-\omega_{r})][\Tilde{\Gamma}_0/2-i(\omega-\omega_q)]+g_c^2}.  \nonumber
\end{equation}

Here, $\Tilde{\Gamma}_0$ is the decoherence rate of the DQD at the interdot charge transfer region, $g_c$ is the cavity-DQD coupling constant, $\omega_q=E_q/\hbar$ is the DQD qubit frequency, and $\kappa = \kappa_c+\kappa_\text{int}$ is the total losses in the cavity. Figure~\ref{fig2}(d) shows the theoretically fitted result with tunnel coupling \textit{t}, cavity-DQD coupling $g_c$, and decoherence rate $\Tilde{\Gamma}_0$ being the fitting parameters. With the fitted values of $g_c$ and $\delta_r$, the dashed lines in Fig.~\ref{fig2}(c) show the lowest two transition energies of the Jaynes-Cummings Hamiltonian, which is given by~\cite{blais2004,ungerer2023, frey2012}:

\begin{equation}\label{Eq:dispersive}
    \omega_\text{JC}= \frac{(\omega_q+\omega_r)}{2}\pm\frac{1}{2}\sqrt{(\omega_q-\omega_r)^2+4g_c^2}.
\end{equation}

The above fitting procedure yields \textit{t} =  $2\:\pm\:1$~$\mu$eV (correspondingly, $\delta_r$ = 27.5~$\mu$eV) and $g_c/2\pi$ = $43\:\pm\:4$~MHz, and $\Gamma_0/2\pi = 2600~\pm~200$~MHz as the decoherence rate when tunneling in and out the DQD is suppressed. In Fig.~\ref{fig2}(c), the spectral response of the cavity in the Coulomb blockade regime ($\delta\gg\delta_r$) provides the input-coupling $\kappa_c/2\pi$ = 3.9~MHz and internal losses of the cavity $\kappa_\text{int}/2\pi$ = 1.3~MHz~\cite{khan2021}. Note that these loss parameters also define the quality factors of the cavity $Q_\text{L}=\omega_r/(\kappa_c+\kappa_\text{int})$ and $Q_\text{ex}=\omega_r/\kappa_\text{c}$.

Moving next, the photocurrent $I_\text{SD}^\pm$ in the low-power linear-response regime is given by~\cite{khan2021, zenelaj2022, haldar2023}:

\begin{equation}\label{eq:lowpower}
    I_\text{SD}/e\dot{N}=\frac{\kappa_\text{c}}{\kappa}\frac{4\kappa_\text{DQD}\kappa}{(\kappa_\text{DQD}+\kappa)^2}\frac{\Gamma_\text{0e}}{(\Gamma_\text{0e}+\gamma_-)}p_f,
\end{equation}

where $\dot{N}=P_\text{in}/\hbar\omega_r$ is the input photon flux, and $p_f$ = $(\Gamma_{g\text{L}}\Gamma_{\text{R}e} -\Gamma_{g\text{R}}\Gamma_{\text{L}e})/\Gamma_{0e}\Gamma_{g0}$ the directivity of the photocurrent between the source (S) and drain (D) reservoirs. Here the tunnel in rate from S to ground state $\ket{g}$ is denoted as $\Gamma_{g\text{L}}=\Gamma_\text{L}\cos^2(\theta/2)$, and tunnel out rate from the excited state $\ket{e}$ to D is $\Gamma_{\text{R}e}= 2\Gamma_\text{R}\cos^2(\theta/2)$, with $\theta=\cos^{-1}(-\delta_r/\hbar\omega_r)$ being the mixing angle. Similarly, for the other two rates, $\Gamma_{g\text{R}}=\Gamma_\text{R}\sin^2(\theta/2)$ and $\Gamma_{\text{L}e}=2\Gamma_\text{L}\sin^2(\theta/2)$. The factor of two appearing for the out-tunneling rates $\Gamma_{\text{R}e}$ and $\Gamma_{\text{L}e}$ accounts for the spin degeneracy of the dot levels~\cite{hofmann2016, barker2022}. Figures~\ref{fig4}(a) and \ref{fig4}(b) schematically illustrate the tunnel in and out rates for the two operation points. The total tunnel rate into $\ket{g}$ (out of $\ket{e}$) is denoted by $\Gamma_{g0}=\Gamma_{g\text{L}}+\Gamma_{g\text{R}}$ ($\Gamma_{0e}=\Gamma_{\text{L}e}+\Gamma_{\text{R}e}$). 

In the strong drive limit, the photocurrent reads~\cite{haldar2023, zenelaj2022}:

\begin{equation}\label{eq:highpower}
    I_{\text{SD}}/e = \frac{16\kappa_{\text{c}}\dot{N}g_c^2(\Gamma_{g\text{L}}\Gamma_{\text{R}e}-\Gamma_{g\text{R}}\Gamma_{\text{L}e})}{16\kappa_{\text{c}}\dot{N}g_c^2(\Gamma_{0e}+2\Gamma_{g0})+\Gamma_{g0}\tilde{\Gamma}\kappa^2(\Gamma_{0e}+\gamma_-)}
\end{equation}

With the already obtained values of $t$, $g_c$, $\kappa_c$ and $\kappa_\text{int}$ using Figs.~\ref{fig2}(c) and \ref{fig2}(d), we here fit $I_\text{SD}^\pm$ in the linear (non-linear) response regime of Fig.~\ref{fig3}(a) using Eq.~(\ref{eq:lowpower}) (Eq.~(\ref{eq:highpower})). The solid (dash-dotted) lines show the theory results in the low (high) power regime, which yields $\Gamma_{\text{L}}/2\pi$ = 12~MHz, $\Gamma_{\text{R}}/2\pi$ = 1100~MHz, inter-dot relaxation rate $\gamma_-/2\pi$ = 23~MHz, and dephasing rate $\gamma_\phi/2\pi$ = 1800~MHz. We find a large $\Gamma_{\text{R}}$ when compared to $\Gamma_{\text{L}}$ and $\gamma_-$, which is causing the asymmetry in the photodiode operation. We also calculate $\kappa_\text{DQD}^+/2\pi$ = 0.8~MHz and $\kappa_\text{DQD}^-/2\pi$ = 1~MHz, where $\kappa_\text{DQD}=4g_c^2/\Tilde{\Gamma}$ and $\Tilde{\Gamma} = \Gamma_{0e}+\gamma_-+4\gamma_\phi$. 

With the above-calculated $\kappa_\text{DQD}^+$ and $\kappa_\text{DQD}^-$ values, the solid red and blue lines in Fig.~\ref{fig2}(f) show the theoretically predicted RF reflectance curves at $\pm\delta_r$ which is given by~\cite{khan2021}:

\begin{equation}\label{refl}
    R(\omega)=\frac{(\kappa_\text{int}+\kappa_\text{DQD}-\kappa_c)^2+4(\omega-\omega_r)^2}{(\kappa_\text{int}+\kappa_\text{DQD}+\kappa_c)^2+4(\omega-\omega_r)^2}.
\end{equation}

The theory results in Fig.~\ref{fig2}(f) are in excellent agreement with experimental data and correctly reproduce the asymmetry of the cavity photon dissipation at the two operation points. 

We noted that the dephasing rate, $ \gamma_\phi$, calculated at a charge triple point using Figs.~\ref{fig3}(a) and \ref{fig2}(f), is about a factor of three higher than that obtained across an inter-dot transition using Figs.~\ref{fig2}(c) and \ref{fig2}(d). This increase in $\gamma_\phi$ at the charge triple point may be attributed to the activation of two-level fluctuators nearby or within the DQD when current flows between the source and drain leads of the DQD. 

\section{Device for near-unity photodetection efficiency}

\begin{figure}[b]
\includegraphics[width=3.1in]{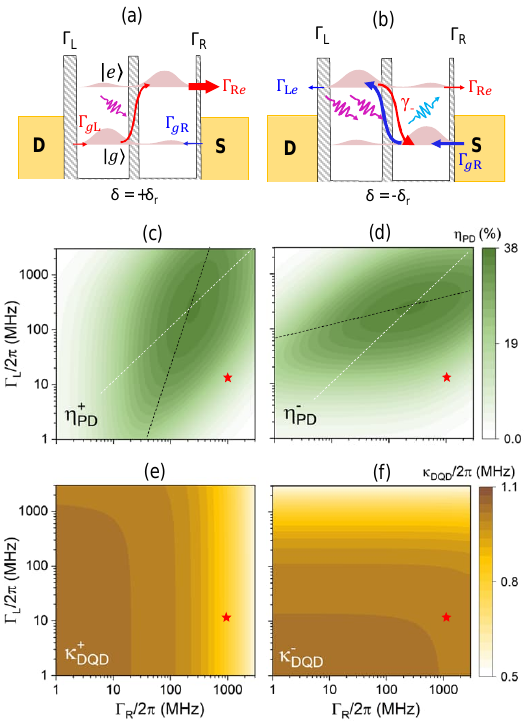}
\caption{\label{fig4}(a-b) Schematic diagrams showing the tunneling in and out rates for the two operation points, highlighting the asymmetry in device operation. Theoretical predictions of quantum efficiency at the low-power limit of Eq.~\ref{eq:lowpower} under varying tunnel couplings $\Gamma_\text{L}$ and $\Gamma_\text{R}$, showing the potential for optimizing photo-conversion efficiency at (c) $+\delta_r$ and (d) $-\delta_r$. (e-f) Show the corresponding cavity photon dissipation due to DQD absorption. The star marks in figures (c-f) show the operation point of the present experiment. The white dashed lines in panels (c) and (d) show the efficiency of a symmetrically tunnel-coupled DQD photodiode. }
\end{figure}

To gain further insights into the device operation and to identify potential improvements needed for near-unity photodetection efficiency, we now analyze each term of Eq.~(\ref{eq:lowpower}) individually~\cite{haldar2023}. The first term, $\kappa_c/\kappa$, and directivity $p_f$ remain constant for the two detuning cases. For the present experiment, we have $\kappa_c/\kappa$ = 0.75 and $p_f$ = 0.67. The second term, $4\kappa_\text{DQD}\kappa/(\kappa_\text{DQD}+\kappa)^2$, describes how well $\kappa_\text{DQD}$ matches the cavity losses $\kappa$. Here, the decoherence, $\Tilde{\Gamma}= (\Gamma_{\text{L}e}+\Gamma_{\text{R}e}) +\gamma_- +4\gamma_\phi$, plays a crucial role in the absorption process in terms of $\kappa_\text{DQD}=4g_c^2/\Tilde{\Gamma}$. For the positive detuning case of Fig.~\ref{fig4}(a), $\Tilde{\Gamma}$ is large because of the larger $\Gamma_\text{Re}$, which leads to a smaller $\kappa_\text{DQD}^+$ and $\eta_\text{PD}^+$ than for the other operation point, Fig.~\ref{fig4}(b). The fractional contribution in $\eta_\text{PD}$ arising from this term is estimated to be 0.46 (0.55) at $+\delta_r$ (-$\delta_r$). Moving to the third term of Eq.~(\ref{eq:lowpower}), $\Gamma_\text{0e}/(\Gamma_\text{0e}+\gamma_-)$, which describes the competition between the tunnelling out ($\Gamma_\text{0e}=\Gamma_{\text{L}e}+\Gamma_{\text{R}e}$) and unwished inter-dot relaxation of a photo-excited electron. With $\Gamma_{\text{R}}$ much larger than $\gamma_-$, in Fig.~\ref{fig4}(a), electrons quickly tunnel out ($\Gamma_{\text{R}e}$) from $\ket{e}$ to S and thus, the effect of relaxation becomes negligible with $+\delta_r$. Whereas, for the other operation point, Fig.~\ref{fig4}(b), with $\Gamma_{\text{L}e}<\gamma_-$ and $\Gamma_{g\text{R}}$, a large number of photo-excited electrons relax to the ground state. Thus, the contribution arising from the third term predicts a larger $\eta_\text{PD}$ with positive detuning, which in our case is 0.99 (0.61) at $+\delta_r$ ($-\delta_r$). 

Thus, the second term showing the mismatch of $\kappa$ and $\kappa_\text{DQD}$ becomes the main reason to observe photodetection efficiency below 50\% level. Additionally, the comparable $\kappa_\text{int}$ and $\kappa_c$ of the cavity, resulting $\kappa_c/\kappa$ = 0.75, causes another $\sim$25\% lowering in $\eta_\text{PD}$. Therefore, a plausible improvement of $\eta_\text{PD}$ would be (i) to employ a high-impedance cavity with a strong cavity-DQD coupling enabling $\kappa_\text{DQD}=4g_c^2/\Gamma=\kappa$~\cite{stockklauser2017, mi2017, ranni2023}, (ii) increasing the input coupling (or, reducing the internal losses) to achieve $\kappa_c \gg \kappa_\text{int}$ and (iii) optimizing the tunnel couplings of the DQD to obtain $p_f= 1$ as described below. Notably, our present photodiode closely aligns with criterion (ii), with $\kappa_c = 3\times\kappa_\text{int}$ and criterion (iii) with directivity $p_f$ = 0.67.

We note that the third term has the biggest difference between the positive and negative detuning contribution and thus defines the asymmetry in photocurrent, with an opposing effect from the second term. To further study how the asymmetry in the DQD influences $\eta_\text{PD}^\pm$, we perform theoretical calculations using Eq.~(\ref{eq:lowpower}) with varying tunnel couplings $\Gamma_\text{L}$ and $\Gamma_\text{R}$ while keeping all other parameters constant, Figs.~\ref{fig4}(c-f). The star marks in the four panels indicate our present device operation point. The dashed white lines in Figs.~\ref{fig4}(c) and \ref{fig4}(d) indicate the symmetric photodiode operation with $\eta_\text{PD}^+=\eta_\text{PD}^-$ and $\kappa_\text{DQD}^+=\kappa_\text{DQD}^-$. The results of Figs.~\ref{fig4}(c) and \ref{fig4}(d) indicate that $\eta_\text{PD}$ of the present device could be increased by a factor of 1.5 by optimizing the tunnel couplings of the DQD. We note that the maximum photodetection efficiency of the device arises with $\Gamma_\text{L} = \Gamma_\text{R}\approx 10\gamma_-$. However, when  $\Gamma_\text{L}$ and $\Gamma_\text{R}$ are much larger or lower than $\sim 10\gamma_-$, the maximum photodetection efficiency can be obtained along an off-diagonal axis, marked with a black dashed line in Figs.~\ref{fig4}(c) and \ref{fig4}(d). The slope of this line depends on the relaxation rate and inter-dot tunnel coupling. In this regime, to achieve a higher $\eta_\text{PD}$ by minimizing the competing DQD relaxation process, it is advantageous to increase one of the tunnel rates at the cost of the other.

For the first-generation device presented in Ref.~\citealp{khan2021}, photon dissipation caused by resonator internal losses, an unnecessary additional port, and photon leakage through the DC gate lines reduced the photodetection efficiency. In this next-generation device, we have added large capacitive filters to the DC lines, minimizing photon leakage through these lines (see Section~\ref{sec2}). Additionally, we replaced the Al resonator used in Ref.~\citealp{khan2021} with an Nb resonator to reduce losses likely caused by impurities and vortices. With these improvements, we achieve $\kappa_\text{int}/2\pi$ = 1.3 MHz, four times lower than the $\kappa_\text{int}/2\pi$ = 5.8 MHz reported in Ref.~\citealp{khan2021}. Furthermore, we have a one-port, over-coupled resonator ($\kappa_c\gg\kappa_\text{int}$), as compared to the two-port under-coupled resonator used in Ref.~\citealp{khan2021}. This allows us to improve the $\eta_\text{PD}$ by increasing the first term of Eq.~(\ref{eq:lowpower}), $\kappa_c/\kappa$, from 0.28 to 0.75. The second, third, and fourth terms in Eq.~(\ref{eq:lowpower}) depend on the tunnel couplings-- $\Gamma_\mathrm{L}$, $\Gamma_\mathrm{R}$, and $t$. We here tune these tunnel couplings to achieve a factor of 1.5 improvement in the third term, describing the competition between the unwanted DQD relaxation $\gamma_-$ and the wished tunneling out rate $\Gamma_{0e}$, as compared to the results presented in Ref.~\citealp{khan2021}. This improvement is obtained at the cost of a slight reduction (from 0.8 to 0.67) in the fourth term that describes the directivity of the DQD.

\section{Performance Parameters}

The $\eta_{\text{PD}}$ calculated as the ratio of the photoelectron tunnel rate $I_{\text{SD}}/e$ to the incident photon flux $P_{\text{in}}/\hbar\omega_r$, accounts cavity photon losses, cavity-DQD coupling and the efficiency of moving an electron from S to D when a photon is absorbed by the DQD. Thus, $\eta_\text{PD}$ corresponds to the external quantum efficiency of the device. 

On the other hand, the internal quantum efficiency $\eta_\text{IQE}$ can be obtained by the ratio of the photon-assisted tunnel rate $I_{\text{SD}}/e$ to the photon absorption rate by the DQD, $R_{\text{abs}} = \kappa_{\text{DQD}} \cdot n_c$. In other words, $\eta_\text{IQE}$ describes how successfully an electron is transferred from S to D once a photon is absorbed by the DQD. Therefore, $\eta_\text{IQE}$ explains the losses due to the interdot relaxation of electrons and electrons lost due to the less-than-unity directivity of the photodiode, $p_f$. In the linear response regime, we obtain $\eta_\text{IQE} = I_\text{SD}/(e\kappa_\text{DQD} n_c)$ = 75\:$\pm$\:8~\% at $+\delta_r$ and $\eta_\text{IQE}$ = 36\:$\pm$\:10~\% at $\delta_{r}$. Remarkably, the DQD, as a photon-to-electron conversion engine at $+\delta_r$, achieves an internal quantum efficiency close to unity. 

Another key performance parameter is the so-called noise equivalent power (NEP), which describes the lowest input power needed to achieve a unity signal-to-noise ratio~\cite{crisci2021}. According to the definition, NEP = $P_\text{min}/\sqrt{B}$, where $B$ is the bandwidth, and $P_\text{min} = \delta I_\text{SD}/\mathbb{R}$ is the lowest input power needed for the unity signal to noise ratio in the detector~\cite{wang2019, guo2024, unal2017}. Here $\delta I_\text{SD}$ is the current noise of the measured signal in the low-power regime with $P_\text{in} \approx P_\text{min}$, and $\mathbb{R}$ the responsivity which we calculate according to $\mathbb{R} = \Delta I_\text{SD}^+/\Delta P_\text{in} = e\eta_\text{PD}/\hbar\omega_r$ = 9000~A/W. To estimate the inherent current noise of our photodetector, we consider shot noise at a diminishing input signal. We observe no leakage current without the applied photon signal in the photodetector operation point within the 40~fA noise level of the current pre-amplifier. Therefore, the dark current is below $I_\text{SD} = 40$~fA. The corresponding shot noise is then less than $\delta I_\text{SD} = \sqrt{2eI_\text{SD} B}$ leading to an upper limit estimate of NEP = 10~zW/$\sqrt{\mathrm{Hz}}$. In our measurements, the current pre-amplifier noise of $\delta I_\text{SD} = 40$~fA with the bandwidth of \textit{B} = 5~Hz, however, dominates over the shot noise. Therefore, in practice, the achieved NEP has a higher value of NEP = $\delta I_\text{SD}/\mathbb{R}\sqrt{B}$ = 2~aW/$\sqrt{\mathrm{Hz}}$. This value is close to the recently reported NEP value of a superconducting microwave detector~\cite{stanisavljevic2023}. As proposed in Ref.~\cite {stanisavljevic2023}, reducing the amplifier noise would improve the operation performance and yield a lower NEP value closer to the inherent limitations of the device.

\vspace{-0.3cm}
\section{Conclusions}
\vspace{-0.2cm}
In conclusion, our present study demonstrated experimental insights into microwave photon-assisted electron tunneling in an asymmetrically tunnel-coupled DQD photodiode. Using a high-quality factor cavity, on-chip filtering to prevent photon leakages, and an optimally chosen operation point to obtain near unity directivity, we achieved a photon-to-electron conversion efficiency of 25\% with an input power level down to 100~aW. The device attains a maximum responsivity of 9000 A/W and a noise equivalent power of 2 aW/$\sqrt{\mathrm{Hz}}$, facilitating efficient photon-to-electron conversion with a single microwave photon in the cavity. Together with a high-speed charge sensing scheme, our work demonstrating continuous high-efficiency microwave photodetection could potentially be useful in solid-state quantum tomography~\cite{izumi2020,pereira2023}, quantum metrology~\cite{matthews2016}, and research in astronomy~\cite{pankratov2022, braggio2025} where the detection of weak microwave signals is of critical need.

\vspace{-0.4cm}
\section*{Acknowledgements}
We acknowledge fruitful discussions with Antti Ranni, Claes Thelander, Adam Burke, Martin Leijnse, and Andreas Wacker and the financial support from the Knut and Alice Wallenberg Foundation through the Wallenberg Center for Quantum Technology (WACQT), the Foundational Questions Institute, a donor advised fund of Silicon Valley Community Foundation (grant number FQXi-IAF19-07), NanoLund, the European Union (ERC, QPHOTON, 101087343), and Swedish Research Council (Dnr 2019-04111). We also thank Jonas Bylander, Simone Gasparinetti, and Anita Fadavi Roudsari from the Chalmers University of Technology for fruitful discussions on improving the superconducting cavity. P.P.P. acknowledges funding from the Swiss National Science Foundation (Eccellenza Professorial Fellowship PCEFP2\_194268).

\bibliography{master}
\end{document}